
%
%
\mathchardef\Delta="7101
\def\footnote#1{\let\@s\empty
  \ifhmode\edef\@s{\spacefactor=\the\spacefactor}\/\fi
  #1\@s\vfootnote{#1}}
\def\m@th{\mathsurround=0pt }
\newdimen\LENB \newdimen\LENW \newdimen\THI
\newdimen\LENWH \newdimen\LENTOT \newcount\N
\def\vbrknlnele#1#2#3{
  \LENB=#1pt \LENW=#2pt \THI=#3pt
  \LENWH=\LENW \divide\LENWH by 2
  \LENTOT=\LENB \advance\LENTOT by \LENW
  \vbox to \LENTOT{
    \vbox to \LENWH{}
    \nointerlineskip
    \vbox to \LENB{\hbox to \THI{\vrule width \THI height \LENB}}
    \nointerlineskip
    \vbox to \LENWH{}}}
\def\vbrknln#1{
  \N=#1
  \vcenter{\vbox{
    \loop\ifnum\N>0
      \vbox to 4pt{\vbrknlnele{2}{2}{0.1}}
      \nointerlineskip
      \advance\N by -1
    \repeat}}}
\def\vbl#1{\hskip-5pt\vbrknln{#1}\hskip-5pt}
\def\hbrknlnele#1#2#3{
  \LENB=#1pt \LENW=#2pt \THI=#3pt
  \LENTOT=\LENB \advance\LENTOT by \LENW
  \vcenter{\vbox to \THI{
    \hbox to \LENTOT{
      \hfil
      \vrule width \LENB height \THI
      \hfil}}}}
\def\hblele{\hbrknlnele{2}{2.2}{0.1}}
\def\hblfil{\cleaders\hbox{$\m@th\mkern1mu\hblele\mkern1mu$}\hfill}
\font\sect=cmssbx10 scaled \magstep1

\magnification=\magstep1
\baselineskip=16pt
\ \par
\ \par
\ \par
\ \par
\centerline{{\sect Casorati Determinant Solution for the Relativistic }}
\par
\centerline{{\sect Toda Lattice Equation}}\par
\ \par
\ \par
\centerline{Yasuhiro OHTA\footnote{$^*$}{On leave from Department of
Applied
            Mathematics, Faculty of Engineering, Hiroshima University.},
            Kenji KAJIWARA$^a$, Junta MATSUKIDAIRA$^b$}\par
\centerline{and Junkichi SATSUMA$^c$}\par
\centerline{\sl Research Institute for Mathematical Sciences,}
\centerline{\sl Kyoto University, Kyoto 606, Japan}
\centerline{$^a${\sl Department of Applied Physics, Faculty of
Engineering,}}
\centerline{\sl University of Tokyo, Tokyo 113, Japan}
\centerline{$^b${\sl Department of Applied Mathematics and
Informatics,}}
\centerline{\sl Ryukoku University, Seta, Ohtsu 520-21, Japan}
\centerline{$^c${\sl Department of Mathematical Sciences, University of
            Tokyo,}}
\centerline{\sl Meguro-ku, Tokyo 153, Japan}
\ \par
\ \par
\ \par
\ \par
\ \par
     The relativistic Toda lattice equation is decomposed into three Toda
systems, the Toda lattice itself, B\"acklund transformation of Toda
lattice
and discrete time Toda lattice.  It is shown that the solutions of the
equation are given in terms of the Casorati determinant.  By using the
Casoratian technique, the bilinear equations of Toda systems are reduced
to the Laplace expansion form for determinants.  The $N$-soliton solution
is explicitly constructed in the form of the Casorati determinant.  \par
\vfill
\eject
\noindent
{\sect\uppercase\expandafter{\romannumeral1}. INTRODUCTION}\par
\ \par
     The relativistic Toda lattice (RT) equation,
$$
 \eqalignno{
  \ddot q_n = &( 1 + {1 \over {\rm c}}\dot q_{n-1} )
               ( 1 + {1 \over {\rm c}}\dot q_n     )
               {\exp(q_{n-1}-q_n) \over
                1 + {1 \over {\rm c}^2}\exp(q_{n-1}-q_n)}\cr
            - &( 1 + {1 \over {\rm c}}\dot q_n     )
               ( 1 + {1 \over {\rm c}}\dot q_{n+1} )
               {\exp(q_n-q_{n+1}) \over
                1 + {1 \over {\rm c}^2}\exp(q_n-q_{n+1})},&(1.1)}
$$
where $q_n$ is the coordinates of $n$-th lattice point, and $\dot{}$ means
the differentiation with respect to time $t$ and ${\rm c}$ is the light
speed, was introduced and studied by Ruijsenaars.{$^{1}$}
This equation
is derived by requiring the Poincar\'e invariance to the well-known
Toda
lattice (TL) equation.  The action-angle transformation and complete
integrability were also investigated for the RT equation.  Bruschi and
Ragnisco{$^{2}$} constructed the recursion operator and B\"acklund
transformation for eq.~(1.1).  The relation between the discrete time
Toda
lattice (DTL) and RT was found by Suris{$^{3}$} in the context of the
Hamiltonian dynamics.  We will show later that the DTL equation
actually
takes a part of the RT equation.  In the limit of ${\rm c}\to\infty$,
the RT
eq.~(1.1) reduces to the ordinary TL equation,
$$
 \ddot q_n = \exp(q_{n-1}-q_n) - \exp(q_n-q_{n+1}),\eqno(1.2)
$$
which is one of the most typical equations in soliton theory.  \par
     The bilinear formalism introduced by Hirota{$^{4}$} is a key to
understand the algebraic structure of soliton equations.
By variable transformations,
soliton equations are transformed into the bilinear form.
The solutions
for the bilinear equations are given in terms of certain determinants
or
Pfaffians.  We analyze the RT equation from the view point of the
bilinear
theory.  The Wronski determinant representation of solutions was first
given for the KdV equation.{$^{5}$}  Freeman and Nimmo{$^{6}$} have
developed the Laplace expansion technique and given the Wronskian
solutions
for various soliton equations.  Hirota, Ito and Kako{$^{7}$} showed that
for the discrete analogues of soliton equations, solutions for their
bilinear
form are given by the Casorati determinant instead of Wronskian.  \par
     M. and Y. Sato{$^{8,9}$} found that the soliton equations are
reduced
to the Pl\"ucker relation with respect to the $\tau$ function which is
expressed by the infinite dimensional Wronski determinant.
Date, Jimbo,
Kashiwara and Miwa{$^{10,11}$} developed the theory of transformation
group
for solutions of soliton equations.  Discrete soliton equations were
also
treated in their context.{$^{11,12}$}  Ueno and Takasaki{$^{13}$} extended
Sato's idea to the Toda lattice hierarchy.  \par
     Recently Hietarinta and one of the authors (J.S.)
found that the RT equation
is transformed into a trilinear form through a suitable
dependent variable transformation.{$^{14}$}  It was also shown that
solutions for the trilinear equation are given by the determinant with
Wronskian structure in two directions, rowwise and columnwise.  The
trilinear equation can be regarded as an extention of the bilinear
form.  Therefore it may be expected that the trilinear form has abundant
algebraic structure.{$^{15}$}  However the trilinear equation for RT
is in the case that the coupling constant is negative, and hence the situation
is not considered to be physical.  To overcome this difficulty, we have
investigated
the solutions of eq.~(1.1) and it has become clear that this equation holds
various algebraic structures simultaneously.  \par
     In this paper, we propose another type of determinant solution to the
RT eq.~(1.1), the Casorati determinant.  This determinant represents the
$N$-soliton solution for RT, so fits for the physical situation.  The
solution converges to the Wronskian solution for the one-dimensional TL
equation in the limit of ${\rm c}\to\infty$.  The RT equation is transformed
into three bilinear equations, the TL equation, the B\"acklund
transformation (BT) for TL and the DTL equation.  We show how the RT equation
is composed of these three Toda systems.  It is proved by using the Laplace
expansion and the reduction technique that the Casorati determinant satisfies
the bilinear equations.  For discrete soliton equations, the difference
formula is simply obtained by using the Casoratian technique.{$^{16}$}
Through the differential formula and difference formula for the Casorati
determinant, the bilinear equations of RT are reduced to the ``algebraic"
identities.  In {\sect\uppercase\expandafter{\romannumeral 2}},
the bilinearization of the RT equation is carried out
by introducing the auxiliary dependent variables.
In {\sect\uppercase\expandafter{\romannumeral 3}}, we show the
Casorati determinant solution for the bilinear equations which represents
the $N$-soliton solution.  A constructive proof that the determinant
satisfies the RT equation is given in {\sect\uppercase\expandafter
{\romannumeral 4}},  and {\sect\uppercase\expandafter{\romannumeral 5}}.
At first in {\sect\uppercase\expandafter{\romannumeral 4}}, the
solutions for the two-dimensional Toda lattice (2DTL) equation, BT of TL
and discrete time two-dimensional Toda lattice (D2DTL) equation are derived
in the form of the Wronski and Casorati determinants.
Next in {\sect\uppercase\expandafter{\romannumeral 5}}, we
explain the reduction technique which reduces the 2DTL and D2DTL equations
into the TL and DTL equations, respectively.  In
{\sect\uppercase\expandafter{\romannumeral 6}}, concluding remarks
are given.  \par
\ \par
\noindent
{\sect\uppercase\expandafter{\romannumeral2}.
BILINEAR FORM FOR THE RT EQUATION}\par
\ \par     The RT eq.~(1.1) is transformed into the three bilinear equations,
$$
 \eqalignno{
  &D_x^2f_n\cdot f_n = 2( {\bar g}_ng_n - f_n^2 ),&(2.1{\rm a}) \cr
  &(aD_x-1)f_n\cdot f_{n-1} + {\bar g}_{n-1}g_n = 0,&(2.1{\rm b}) \cr
  &{\bar g}_{n-1}g_{n+1} - f_n^2 = a^2( f_{n+1}f_{n-1} - f_n^2 ),
   &(2.1{\rm c})}
$$
through the variable transformations,
$$
 \eqalignno{
  &q_n = \log{f_{n-1} \over f_n},&(2.2{\rm a}) \cr
  &t = {\sqrt{1+{\rm c}^2} \over {\rm c}}x,&(2.2{\rm b}) \cr
  &{\rm c} = {\sqrt{1-a^2} \over a}.&(2.2{\rm c})}
$$
In eqs.~(2.1), $g$ and $\bar g$ are the auxiliary variables and $D$ is
the bilinear differential operator defined by{$^{4}$}
$$
 D_x^k f\cdot g = (\partial_x-\partial_y)^k f(x)g(y)|_{y=x}.\eqno(2.3)
$$
As will be shown later, eqs.~(2.1a), (2.1b) and (2.1c) are nothing but
the bilinear forms of TL equation, BT of TL and DTL equation,
respectively.  \par
     Let us show that the RT equation actually follows from the bilinear
eqs.~(2.1).  Dividing eqs.~(2.1a), (2.1b) and (2.1c) by $2f_n^2$,
$f_{n-1}f_n$ and $f_{n-1}f_{n+1}$ respectively, we get
$$
 \eqalignno{
  &{{\bar g}_ng_n \over f_nf_n} = 1 + (\log f_n)_{xx},&(2.4{\rm a}) \cr
  &{{\bar g}_{n-1}g_n \over f_{n-1}f_n} = 1 + a(\log {f_{n-1} \over f_n})_x,
   &(2.4{\rm b}) \cr
  &{{\bar g}_{n-1}g_{n+1} \over f_{n-1}f_{n+1}}
   = a^2 + (1-a^2){f_nf_n \over f_{n-1}f_{n+1}},&(2.4{\rm c})}
$$
where the subscript $x$ indicates differentiation.  We have an identity
$$
 {{\bar g}_{n-1}g_{n+1} \over f_{n-1}f_{n+1}}{{\bar g}_ng_n \over f_nf_n}
 = {{\bar g}_{n-1}g_n \over f_{n-1}f_n}{{\bar g}_ng_{n+1} \over f_nf_{n+1}},
 \eqno(2.5)
$$
which is rewritten by using eqs.~(2.4) as
$$
 ( a^2 + (1-a^2){f_nf_n \over f_{n-1}f_{n+1}} )( 1 + (\log f_n)_{xx} )
 = ( 1 + a(\log {f_{n-1} \over f_n})_x )( 1 + a(\log {f_n \over f_{n+1}})_x ).
 \eqno(2.6)
$$
Noticing
$$
 {1-a^2 \over a^2} = {\rm c}^2,\eqno(2.7)
$$
$$
 \partial_x = {\sqrt{1+{\rm c}^2} \over {\rm c}}\partial_t
 = {1 \over a{\rm c}}\partial_t,\eqno(2.8)
$$
we get from eq.~(2.6)
$$
 a^2 + {1 \over {\rm c}^2}(\log f_n)_{tt}
 = ( 1 + {1 \over {\rm c}}\dot q_n )( 1 + {1 \over {\rm c}}\dot q_{n+1} )
  /( 1 + {\rm c}^2\exp({q_{n+1}-q_n}) ).\eqno(2.9)
$$
Finally subtracting eq.~(2.9) from eq.~(2.9) with $n$ replaced by $n-1$,
we obtain
$$\eqalignno{
 {1 \over {\rm c}^2}\ddot q_n
& = ( 1 + {1 \over {\rm c}}\dot q_{n-1} )( 1 + {1 \over {\rm c}}\dot q_n )
  /( 1 + {\rm c}^2\exp({q_n-q_{n-1}}) )\cr
& - ( 1 + {1 \over {\rm c}}\dot q_n )( 1 + {1 \over {\rm c}}\dot q_{n+1} )
  /( 1 + {\rm c}^2\exp({q_{n+1}-q_n}) ),&(2.10)\cr}
$$
which is the RT eq.~(1.1) itself.  Therefore if $f_n$, $g_n$ and ${\bar g}_n$
satisfy eqs.~(2.1), $q_n$ defined by eq.~(2.2a) gives the solution for RT
equation.  \par
\ \par
\noindent
{\sect\uppercase\expandafter{\romannumeral3}. CASORATI DETERMINANT
SOLUTION FOR THE RT EQUATION}\par
\ \par
     The Casorati determinant representation of soliton solution for the
bilinear RT eq.\par
\noindent (2.1) is given by
$$
 f_n = \left|\matrix{
  \varphi_1^{(n)} &\varphi_1^{(n+1)} &\cdots &\varphi_1^{(n+N-1)} \cr
  \varphi_2^{(n)} &\varphi_2^{(n+1)} &\cdots &\varphi_2^{(n+N-1)} \cr
  \vdots          &\vdots            &       &\vdots              \cr
  \varphi_N^{(n)} &\varphi_N^{(n+1)} &\cdots &\varphi_N^{(n+N-1)} \cr}
 \right|,\eqno(3.1{\rm a})
$$
$$
 g_n = \left|\matrix{
  \psi_1^{(n)} &\psi_1^{(n+1)} &\cdots &\psi_1^{(n+N-1)} \cr
  \psi_2^{(n)} &\psi_2^{(n+1)} &\cdots &\psi_2^{(n+N-1)} \cr
  \vdots       &\vdots         &       &\vdots           \cr
  \psi_N^{(n)} &\psi_N^{(n+1)} &\cdots &\psi_N^{(n+N-1)} \cr}
 \right|,\eqno(3.1{\rm b})
$$
$$
 {\bar g}_n = \left|\matrix{
  \bar\psi_1^{(n)} &\bar\psi_1^{(n+1)} &\cdots &\bar\psi_1^{(n+N-1)} \cr
  \bar\psi_2^{(n)} &\bar\psi_2^{(n+1)} &\cdots &\bar\psi_2^{(n+N-1)} \cr
  \vdots           &\vdots             &       &\vdots               \cr
  \bar\psi_N^{(n)} &\bar\psi_N^{(n+1)} &\cdots &\bar\psi_N^{(n+N-1)} \cr}
 \right|,\eqno(3.1{\rm c})
$$
and
$$
 \varphi_i^{(n)} = ({p_i \over 1-p_ia})^n{\rm e}^{\eta_i}
  + ({p_i^{-1} \over 1-p_i^{-1}a})^n{\rm e}^{\xi_i},\eqno(3.2{\rm a})
$$
$$
 \psi_i^{(n)} = p_i^{-1}({p_i \over 1-p_ia})^n{\rm e}^{\eta_i}
  + p_i({p_i^{-1} \over 1-p_i^{-1}a})^n{\rm e}^{\xi_i},\eqno(3.2{\rm b})
$$
$$
 \bar\psi_i^{(n)} = p_i({p_i \over 1-p_ia})^n{\rm e}^{\eta_i}
  + p_i^{-1}({p_i^{-1} \over 1-p_i^{-1}a})^n{\rm e}^{\xi_i},\eqno(3.2{\rm c})
$$
$$
 \eta_i = p_ix + \eta_{i0},\eqno(3.3{\rm a})
$$
$$
 \xi_i = p_i^{-1}x + \xi_{i0},\eqno(3.3{\rm b})
$$
where $N$ is the number of solitons, and $p_i$ and
$\eta_{i0}-\xi_{i0}$ are
the wave number and phase parameter of $i$-th soliton respectively.  \par
     In the non-relativistic limit, ${\rm c}$ tends to infinity, $a$ becomes
$0$ and
$$
 g_{n+1} = {\bar g}_{n-1} = f_n,\eqno(3.4)
$$
$$
 x = t.\eqno(3.5)
$$
In this case, eq.~(2.1a) becomes the bilinear form of TL equation,
$$
 D_t^2f_n\cdot f_n = 2( f_{n+1}f_{n-1} - f_n^2 ),\eqno(3.6)
$$
and eqs.~(2.1b) and (2.1c) are trivial.  Equation (3.6) is transformed
into the TL eq.~(1.2) through the variable transformation (2.2a).  At
${\rm c}\to\infty$, the determinant $f_n$ in eq.~(3.1a) gives the $N$-soliton
solution to the TL equation in Wronskian form.  \par
     We will prove by using the Wronskian technique and Casoratian technique
that the bilinear eqs.~(2.1) are satisfied by $f_n$, $g_n$ and ${\bar g}_n$
in eqs.~(3.1).  At first we give the determinant solution for the 2DTL,
BT of TL and D2DTL equations in {\sect\uppercase\expandafter{\romannumeral 4}}.
 Next in {\sect\uppercase\expandafter{\romannumeral 5}},
applying the reduction technique we recover the Casorati determinant solution
in eqs.~(3.1).  \par
\ \par
\noindent
{{\sect\uppercase\expandafter{\romannumeral4}. THREE TODA SYSTEMS AND THEIR
DETERMINANT SOLUTIONS}\par
\ \par
     In this section we show that the solutions for the 2DTL equation,
BT of TL and D2DTL equation are given in terms of the Wronskian, Casorati
determinant and also Casorati determinant, respectively.  \par
\ \par
\noindent
{\sect A. 2DTL}\par
\ \par
     The bilinear form of 2DTL equation is written as
$$
 D_{x_1}D_{x_{-1}}\tau_n\cdot\tau_n
  = 2( \tau_n\tau_n - \tau_{n+1}\tau_{n-1} ).\eqno(4.1)
$$
The Wronskian solution for this equation is given by{$^{7,17}$}
$$
 \tau_n = \left|\matrix{
  \varphi_1^{(n)} &\varphi_1^{(n+1)} &\cdots &\varphi_1^{(n+N-1)} \cr
  \varphi_2^{(n)} &\varphi_2^{(n+1)} &\cdots &\varphi_2^{(n+N-1)} \cr
  \vdots          &\vdots            &       &\vdots              \cr
  \varphi_N^{(n)} &\varphi_N^{(n+1)} &\cdots &\varphi_N^{(n+N-1)} \cr}
 \right|,\eqno(4.2)
$$
where $\varphi_i^{(n)}$'s are arbitrary functions satisfying the dispersion
relations,
$$
 \partial_{x_\nu}\varphi_i^{(n)} = \varphi_i^{(n+\nu)},\quad \nu=1,-1.
 \eqno(4.3)
$$
After Freeman and Nimmo,{$^{6}$} we use the following notation for
simplicity,
$$
 |n_1,n_2,\cdots,n_N| = \left|\matrix{
  \varphi_1^{(n_1)} &\varphi_1^{(n_2)} &\cdots &\varphi_1^{(n_N)} \cr
  \varphi_2^{(n_1)} &\varphi_2^{(n_2)} &\cdots &\varphi_2^{(n_N)} \cr
  \vdots            &\vdots            &       &\vdots            \cr
  \varphi_N^{(n_1)} &\varphi_N^{(n_2)} &\cdots &\varphi_N^{(n_N)} \cr}
 \right|.\eqno(4.4)
$$
In this notation, $\tau_n$ in eq.~(4.2) is rewritten as
$$
 \tau_n = |n,n+1,\cdots,n+N-1|.\eqno(4.5)
$$
\par
     We show that the above Wronskian satisfies the bilinear eq.~(4.1)
by using the Wronskian technique.  From eq.~(4.3), the differentiation of
the determinant $|n_1,n_2,\cdots,n_N|$ is given by
$$
 \partial_{x_\nu}|n_1,n_2,\cdots,n_N|
  = \sum_{j=1}^N |n_1,n_2,\cdots,n_j+\nu,\cdots,n_N|,\quad \nu=1,-1.\eqno(4.6)
$$
Thus we get the following differential formula for $\tau_n$,
$$
 \eqalignno{
  \partial_{x_1}\tau_n &= |n,n+1,\cdots,n+N-2,n+N|,&(4.7{\rm a}) \cr
  \partial_{x_{-1}}\tau_n &= |n-1,n+1,\cdots,n+N-2,n+N-1|,&(4.7{\rm b}) \cr
  \partial_{x_1}\partial_{x_{-1}}\tau_n
   &= |n-1,n+1,\cdots,n+N-2,n+N| \cr
   &+ |n,n+1,\cdots,n+N-2,n+N-1|.&(4.7{\rm c})}
$$
\par
     Now we consider an identity for $2N\times 2N$ determinant,
$$
 \left|\matrix{
  {}_{n-1} &\vbl4 &{}_{n+1} &{}_{\cdots} &{}_{n+N-2} &{}_{n+N} &\vbl4 &{}_{n}
 &\vbl4 &    &\hbox{\O}     &      &\vbl4 &{}_{n+N-1} \cr
 \multispan{14}\hblfil \cr
  {}_{n-1} &\vbl4 &    &\hbox{\O} &      &    &\vbl4 &{}_{n} &\vbl4 &{}_{n+1}
 &\cdots      &{}_{n+N-2} &\vbl4 &{}_{n+N-1} \cr}
 \right| = 0,\eqno(4.8)
$$
where a number denotes a column vector of height $N$ and \O\ means an
empty block.  Applying the Laplace expansion for determinant to the
left-hand side of above equation, we obtain the algebraic bilinear identity
for determinants,
$$
 \eqalignno{
    &|n-1,n+1,\cdots,n+N-2,n+N|~|n,n+1,\cdots,n+N-2,n+N-1| \cr
  - &|n,n+1,\cdots,n+N-2,n+N|~|n-1,n+1,\cdots,n+N-2,n+N-1| \cr
  + &|n+1,\cdots,n+N-2,n+N-1,n+N|~|n-1,n,n+1,\cdots,n+N-2| = 0.&(4.9)}
$$
By using eqs.~(4.7), this equation is rewritten into the differential
bilinear equation for $\tau_n$,
$$
   (\partial_{x_1}\partial_{x_{-1}}\tau_n-\tau_n)\tau_n
 - \partial_{x_1}\tau_n\partial_{x_{-1}}\tau_n
 + \tau_{n+1}\tau_{n-1} = 0,\eqno(4.10)
$$
which recovers eq.~(4.1).  Hence we have proved that the Wronskian $\tau_n$
gives the solution for 2DTL equation.  \par
\ \par
\noindent
{\sect B. BT of TL}\par
\ \par
     We have the bilinear form for BT of TL equation,{$^{18}$}
$$
 ( aD_x - 1 )\tau_n'\cdot\tau_{n-1} + \tau_n\tau_{n-1}' = 0,\eqno(4.11)
$$
where $a$ is a constant and $x$ corresponds to $x_1$ in eq.~(4.1), which
transforms a TL $\tau_n$ into another TL $\tau_n'$.  Applying the BT
recursively, we get an infinite sequence of TL's, $\tau_n$, $\tau_n'$,
$\tau_n''$, $\cdots$.  We denote the $k$-th TL as $\tau_n(k)$.  Then the
BT of TL is written as
$$
 ( aD_x - 1 )\tau_n(k+1)\cdot\tau_{n-1}(k) + \tau_n(k)\tau_{n-1}(k+1) = 0.
 \eqno(4.12)
$$
\par
     The Casorati determinant solution for eq.~(4.12) is given by
$$
 \tau_n(k) = \left|\matrix{
  \varphi_1^{(n)}(k) &\varphi_1^{(n+1)}(k) &\cdots &\varphi_1^{(n+N-1)}(k) \cr
  \varphi_2^{(n)}(k) &\varphi_2^{(n+1)}(k) &\cdots &\varphi_2^{(n+N-1)}(k) \cr
  \vdots             &\vdots               &       &\vdots                 \cr
  \varphi_N^{(n)}(k) &\varphi_N^{(n+1)}(k) &\cdots &\varphi_N^{(n+N-1)}(k) \cr}
 \right|,\eqno(4.13)
$$
where $\varphi_i^{(n)}(k)$'s are arbitrary functions satisfying the
dispersion relations,
$$
 \partial_x\varphi_i^{(n)}(k) = \varphi_i^{(n+1)}(k),\eqno(4.14{\rm a})
$$
$$
 \Delta_k\varphi_i^{(n)}(k) = \varphi_i^{(n+1)}(k),\eqno(4.14{\rm b})
$$
where $\Delta_k$ is the backward difference operator with the difference
interval $a$ defined by
$$
 \Delta_k f(k) = {f(k)-f(k-1) \over a}.\eqno(4.15)
$$
For simplicity we introduce a convenient notation,
$$
 |{n_1}_{k_1},{n_2}_{k_2},\cdots,{n_N}_{k_N}| = \left|\matrix{
  \varphi_1^{(n_1)}(k_1) &\varphi_1^{(n_2)}(k_2) &\cdots
   &\varphi_1^{(n_N)}(k_N) \cr
  \varphi_2^{(n_1)}(k_1) &\varphi_2^{(n_2)}(k_2) &\cdots
   &\varphi_2^{(n_N)}(k_N) \cr
  \vdots                 &\vdots                 &
   &\vdots                 \cr
  \varphi_N^{(n_1)}(k_1) &\varphi_N^{(n_2)}(k_2) &\cdots
   &\varphi_N^{(n_N)}(k_N) \cr}
 \right|.\eqno(4.16)
$$
In this notation, the solution for BT of TL, $\tau_n(k)$ in eq.~(4.13),
is rewritten as
$$
 \tau_n(k) = |n_k,n+1_k,\cdots,n+N-1_k|,\eqno(4.17)
$$
or suppressing the index $k$,
$$
 \tau_n(k) = |n,n+1,\cdots,n+N-1|.\eqno(4.18)
$$
\par
     We show that the above $\tau_n(k)$ actually satisfies the bilinear
eq.~(4.12).  At first we investigate the difference formula for $\tau$.
{}From eq.~(4.14b) we have
$$
 a\varphi_i^{(n+1)}(k+1) = \varphi_i^{(n)}(k+1) - \varphi_i^{(n)}(k).
  \eqno(4.19)
$$
Noticing this relation, we get
$$
 \eqalignno{
  \tau_n(k+1) &= |n_{k+1},n+1_{k+1},n+2_{k+1},\cdots,n+N-1_{k+1}| \cr
              &= |n_k,n+1_{k+1},n+2_{k+1},\cdots,n+N-1_{k+1}| \cr
\noalign{\hbox{where we have subtracted the 2nd column multiplied by $a$
from the 1st column, }}
              &= |n_k,n+1_k,n+2_{k+1},\cdots,n+N-1_{k+1}| \cr
\noalign{\hbox{where we have subtracted the 3rd column multiplied by $a$
from the 2nd column, }}
              &\cdots \cr
              &= |n_k,n+1_k,\cdots,n+N-2_k,n+N-1_{k+1}|,}
$$
that is,
$$
 \tau_n(k+1) = |n_k,n+1_k,\cdots,n+N-2_k,n+N-1_{k+1}|,\eqno(4.20)
$$
or suppressing the index $k$,
$$
 \tau_n(k+1) = |n,n+1,\cdots,n+N-2,n+N-1_{k+1}|.\eqno(4.21)
$$
Moreover in eq.~(4.20), multiplying the $N$-th column by $a$ and adding
the $(N-1)$-th column to the $N$-th column, we get
$$
 a\tau_n(k+1) = |n_k,n+1_k,\cdots,n+N-2_k,n+N-2_{k+1}|.\eqno(4.22)
$$
The important point is that the effect of difference appears only one column
as shown in eqs.~(4.20) and (4.22).  Consequently the Laplace expansion
becomes applicable.  This technique for condensing the shift of index into
only one column may be called a Casoratian technique.  \par
     Differentiating eq.~(4.22) with $x$ and using eq.~(4.14a) we obtain
$$
 \eqalignno{
  a\partial_x\tau_n(k+1)
   &= |n_k,n+1_k,\cdots,n+N-3_k,n+N-1_k,n+N-2_{k+1}| \cr
   &+ |n_k,n+1_k,\cdots,n+N-3_k,n+N-2_k,n+N-1_{k+1}| \cr
   &= |n_k,n+1_k,\cdots,n+N-3_k,n+N-1_k,n+N-2_{k+1}| \cr
   &+ \tau_n(k+1),&(4.23)}
$$
where we have used eq.~(4.20) in the last line.  We have also
$$
 \partial_x\tau_{n-1}(k) = |n-1,n,\cdots,n+N-3,n+N-1|.\eqno(4.24)
$$
\par
     Let us introduce an identity for $2N\times 2N$ determinant,
$$
 \left|\matrix{
  {}_{n_k} &{}_{\cdots}    &{}_{n+N-3_k} &\vbl4 &{}_{n+N-1_k} &\vbl4
&{}_{n+N-2_{k+1}} &\vbl4
   &      &    &\hbox{\O} &        &\vbl4 &{}_{n+N-2_k} \cr
 \multispan{14}\hblfil \cr
      &\hbox{\O} &        &\vbl4 &{}_{n+N-1_k} &\vbl4 &{}_{n+N-2_{k+1}} &\vbl4
   &{}_{n-1_k} &{}_{n_k} &\cdots    &{}_{n+N-3_k} &\vbl4 &{}_{n+N-2_k} \cr}
 \right| = 0.\eqno(4.25)
$$
Applying the Laplace expansion to the left-hand side, we obtain the
algebraic bilinear identity for determinants,
$$
 \eqalignno{
    &|{}_{n_k},{}_{\cdots},{}_{n+N-3_k},{}_{n+N-1_k},{}_{n+N-2_{k+1}}|
     |{}_{n-1_k},{}_{n_k},{}_{\cdots},{}_{n+N-3_k},{}_{n+N-2_k}| \cr
  - &|{}_{n_k},{}_{\cdots},{}_{n+N-3_k},{}_{n+N-2_k},{}_{n+N-2_{k+1}}|
     |{}_{n-1_k},{}_{n_k},{}_{\cdots},{}_{n+N-3_k},{}_{n+N-1_k}| \cr
  - &|{}_{n_k},{}_{\cdots},{}_{n+N-3_k},{}_{n+N-1_k},{}_{n+N-2_k}|
     |{}_{n-1_k},{}_{n_k},{}_{\cdots},{}_{n+N-3_k},{}_{n+N-2_{k+1}}|
= 0,&(4.26)}
$$
which is rewritten by using eqs.~(4.20), (4.22), (4.23) and (4.24), into
the differential bilinear equation,
$$
   (a\partial_x\tau_n(k+1)-\tau_n(k+1))\tau_{n-1}(k)
 - a\tau_n(k+1)\partial_x\tau_{n-1}(k)
 + \tau_n(k)\tau_{n-1}(k+1) = 0.\eqno(4.27)
$$
This equation is equal to eq.~(4.12).  Hence we have proved that the Casorati
determinant $\tau_n(k)$ in eq.~(4.13) gives the solution for the BT of TL
equation.  \par
\ \par
\noindent
{\sect C. D2DTL}\par
\ \par
     Using the Casoratian technique, we show that the D2DTL equation is
satisfied by the Casorati determinant.  The bilinear form of D2DTL equation
is given by{$^{7,19}$}
$$\eqalignno{
& \tau_n(k+1,l+1)\tau_n(k,l) - \tau_n(k+1,l)\tau_n(k,l+1)\cr
= & ab( \tau_n(k+1,l+1)\tau_n(k,l) - \tau_{n+1}(k+1,l)\tau_{n-1}(k,l+1) ),
 &(4.28)\cr}
$$
where $\tau_n$ depends on two discrete independent variables $k$ and $l$, and
$a$ and $b$ are the difference intervals for $k$ and $l$ respectively.  \par
     The Casorati determinant solution for eq.~(4.28) is given as
$$
 \tau_n(k,l) = \left|\matrix{
  \varphi_1^{(n)}(k,l) &\varphi_1^{(n+1)}(k,l) &\cdots
   &\varphi_1^{(n+N-1)}(k,l) \cr
  \varphi_2^{(n)}(k,l) &\varphi_2^{(n+1)}(k,l) &\cdots
   &\varphi_2^{(n+N-1)}(k,l) \cr
  \vdots               &\vdots                 &
   &\vdots                 \cr
  \varphi_N^{(n)}(k,l) &\varphi_N^{(n+1)}(k,l) &\cdots
   &\varphi_N^{(n+N-1)}(k,l) \cr}
 \right|,\eqno(4.29)
$$
where $\varphi_i^{(n)}$'s are arbitrary functions of $k$ and $l$ which
satisfy the dispersion relations,
$$ \Delta_k\varphi_i^{(n)}(k,l) = \varphi_i^{(n+1)}(k,l),\eqno(4.30{\rm a})
$$
$$
 \Delta_l\varphi_i^{(n)}(k,l) = \varphi_i^{(n-1)}(k,l).\eqno(4.30{\rm b})
$$
Here $\Delta_k$ and $\Delta_l$ are the backward difference operators defined
by
$$
 \Delta_k f(k,l) = {f(k,l)-f(k-1,l) \over a},\eqno(4.31{\rm a})
$$
$$
 \Delta_l f(k,l) = {f(k,l)-f(k,l-1) \over b}.\eqno(4.31{\rm b})
$$
We use the following simple notation similar to that in eq.~(4.16),
$$
 |{n_1}_{\matrix{{\scriptstyle k_1} \cr
                 \noalign{\vskip-5pt}
                 {\scriptstyle l_1} \cr}},
  {n_2}_{\matrix{{\scriptstyle k_2} \cr
                 \noalign{\vskip-5pt}
                 {\scriptstyle l_2} \cr}},
  \cdots,
  {n_N}_{\matrix{{\scriptstyle k_N} \cr
                 \noalign{\vskip-5pt}
                 {\scriptstyle l_N} \cr}}|
 = \left|\matrix{
  \varphi_1^{(n_1)}(k_1,l_1) &\varphi_1^{(n_2)}(k_2,l_2) &\cdots
   &\varphi_1^{(n_N)}(k_N,l_N) \cr
  \varphi_2^{(n_1)}(k_1,l_1) &\varphi_2^{(n_2)}(k_2,l_2) &\cdots
   &\varphi_2^{(n_N)}(k_N,l_N) \cr
  \vdots                     &\vdots                     &
   &\vdots                     \cr
  \varphi_N^{(n_1)}(k_1,l_1) &\varphi_N^{(n_2)}(k_2,l_2) &\cdots
   &\varphi_N^{(n_N)}(k_N,l_N) \cr}
 \right|.\eqno(4.32)
$$
In this notation, the solution to D2DTL equation in eq.~(4.29) is
$$
 \tau_n(k,l) =
 |n_{\matrix{{\scriptstyle k} \cr
             \noalign{\vskip-5pt}
             {\scriptstyle l} \cr}},
  n+1_{\matrix{{\scriptstyle k} \cr
               \noalign{\vskip-5pt}
               {\scriptstyle l} \cr}},
  \cdots,
  n+N-1_{\matrix{{\scriptstyle k} \cr
                 \noalign{\vskip-5pt}
                 {\scriptstyle l} \cr}}|.\eqno(4.33)
$$
Neglecting the index $k$ and $l$, we write
$$
 \tau_n(k,l) = |n,n+1,\cdots,n+N-1|.\eqno(4.34)
$$
\par
     Now let us examine the difference formula for the above $\tau_n$.
Note that the dispersion relation for $k$ in eq.~(4.31a) is equal to the one
for the solution to BT of TL in {\sect B}.  Hence in the same manner as
shown in eqs.~(4.20) and (4.22), we obtain
$$
 \tau_n(k+1,l) =
 |n_{\matrix{{\scriptstyle k} \cr
             \noalign{\vskip-5pt}
             {\scriptstyle l} \cr}},
  n+1_{\matrix{{\scriptstyle k} \cr
               \noalign{\vskip-5pt}
               {\scriptstyle l} \cr}},
  \cdots,
  n+N-2_{\matrix{{\scriptstyle k} \cr
                 \noalign{\vskip-5pt}
                 {\scriptstyle l} \cr}},
  n+N-1_{\matrix{{\scriptstyle k+1} \cr
                 \noalign{\vskip-5pt}
                 {\scriptstyle l} \cr}}|,\eqno(4.35)
$$
$$
 a\tau_n(k+1,l) =
 |n_{\matrix{{\scriptstyle k} \cr
             \noalign{\vskip-5pt}
             {\scriptstyle l} \cr}},
  n+1_{\matrix{{\scriptstyle k} \cr
               \noalign{\vskip-5pt}
               {\scriptstyle l} \cr}},
  \cdots,
  n+N-2_{\matrix{{\scriptstyle k} \cr
                 \noalign{\vskip-5pt}
                 {\scriptstyle l} \cr}},
  n+N-2_{\matrix{{\scriptstyle k+1} \cr
                 \noalign{\vskip-5pt}
                 {\scriptstyle l} \cr}}|.\eqno(4.36)
$$
Similarly we get
$$
 \tau_n(k,l+1) =
 |n_{\matrix{{\scriptstyle k} \cr
             \noalign{\vskip-5pt}
             {\scriptstyle l+1} \cr}},
  n+1_{\matrix{{\scriptstyle k} \cr
               \noalign{\vskip-5pt}
               {\scriptstyle l} \cr}},
  \cdots,
  n+N-2_{\matrix{{\scriptstyle k} \cr
                 \noalign{\vskip-5pt}
                 {\scriptstyle l} \cr}},
  n+N-1_{\matrix{{\scriptstyle k} \cr
                 \noalign{\vskip-5pt}
                 {\scriptstyle l} \cr}}|,\eqno(4.37)
$$
$$
 b\tau_n(k,l+1) =
 |n+1_{\matrix{{\scriptstyle k} \cr
               \noalign{\vskip-5pt}
               {\scriptstyle l+1} \cr}},
  n+1_{\matrix{{\scriptstyle k} \cr
               \noalign{\vskip-5pt}
               {\scriptstyle l} \cr}},
  \cdots,
  n+N-2_{\matrix{{\scriptstyle k} \cr
                 \noalign{\vskip-5pt}
                 {\scriptstyle l} \cr}},
  n+N-1_{\matrix{{\scriptstyle k} \cr
                 \noalign{\vskip-5pt}
                 {\scriptstyle l} \cr}}|,\eqno(4.38)
$$
because the role of $l$ and $b$ is parallel to that of $k$ and $a$ except
that the ordering of index $n$ is reversed.  In short, we write
$$
 \tau_n(k+1,l) = |n,n+1,\cdots,n+N-2,n+N-1_{k+1}|,\eqno(4.39)
$$
$$
 a\tau_n(k+1,l) = |n,n+1,\cdots,n+N-2,n+N-2_{k+1}|,\eqno(4.40)
$$
$$
 \tau_n(k,l+1) = |n_{l+1},n+1,\cdots,n+N-2,n+N-1|,\eqno(4.41)
$$
$$
 b\tau_n(k,l+1) = |n+1_{l+1},n+1,\cdots,n+N-2,n+N-1|.\eqno(4.42)
$$
\par
     We give the Casoratian technique for two variable shifted case.  From
eq.~(4.41), $\tau_n(k+1,l+1)$ is given by
$$
 \tau_n(k+1,l+1) =
 |n_{\matrix{{\scriptstyle k+1} \cr
             \noalign{\vskip-5pt}
             {\scriptstyle l+1} \cr}},
  n+1_{k+1},\cdots,n+N-2_{k+1},n+N-1_{k+1}|.\eqno(4.43)
$$
In the following we show that the shifts of index $k$ is condensed into
only the most right column of the determinant.
{}From eqs.~(4.30), $\varphi_i^{(n)}$
satisfies
$$
 \Delta_k\Delta_l\varphi_i^{(n)} = \varphi_i^{(n)},\eqno(4.44)
$$
that is,
$$
 \varphi_i^{(n)}(k,l) - \varphi_i^{(n)}(k,l-1) - \varphi_i^{(n)}(k-1,l)
  + \varphi_i^{(n)}(k-1,l-1) = ab\varphi_i^{(n)}(k,l),\eqno(4.45)
$$
which is rewritten as
$$
 \eqalignno{
    (1-ab)\varphi_i^{(n)}(k+1,l+1)
 &= \varphi_i^{(n)}(k+1,l) + \varphi_i^{(n)}(k,l+1) - \varphi_i^{(n)}(k,l) \cr
 &= \varphi_i^{(n)}(k,l+1) + a\varphi_i^{(n+1)}(k+1,l).&(4.46)}
$$
Therefore multiplying the both hand sides of eq.~(4.43) by $(1-ab)$ and
rewriting the 1st column of the determinant by the use of eq.~(4.46),
we obtain
$$
 \eqalignno{
&  (1-ab)\tau_n(k+1,l+1)\cr
  &= |n_{\matrix{{\scriptstyle k} \cr
                 \noalign{\vskip-5pt}
                 {\scriptstyle l+1} \cr}},
      n+1_{k+1},\cdots,n+N-2_{k+1},n+N-1_{k+1}| \cr
  &+ a|n+1_{\matrix{{\scriptstyle k+1} \cr
                    \noalign{\vskip-5pt}
                    {\scriptstyle l} \cr}},
       n+1_{k+1},\cdots,n+N-2_{k+1},n+N-1_{k+1}|.&(4.47)}
$$
The second term in r.h.s. vanishes because its 1st and 2nd columns are
the same.  So we get
$$
 (1-ab)\tau_n(k+1,l+1) = |n_{l+1},n+1_{k+1},\cdots,n+N-2_{k+1},n+N-1_{k+1}|.
 \eqno(4.48)
$$
In the above determinant, by subtracting the $(j+1)$-th column multiplied by
$a$ from the $j$-th column for $j=2,3,\cdots,N-1$, $\tau_n(k+1,l+1)$
is given by
$$
 (1-ab)\tau_n(k+1,l+1) = |n_{l+1},n+1,\cdots,n+N-2,n+N-1_{k+1}|.\eqno(4.49)
$$
As is shown, even if the two variables $k$ and $l$ are shifted, $\tau_n$
is also expressed in the determinant form whose columns are almost unchanged
and only edges are varied.  Thus we can use the Laplace expansion
technique.  \par
     We consider an identity for $2N\times 2N$ determinant,
$$
 \left|\matrix{
  {}_{n_{l+1}} &\vbl4 &{}_{n+1} &{}_{\cdots}    &{}_{n+N-2} &{}_{n+N-1_{k+1}}
 &\vbl4 &{}_n &\vbl4
   &    &\hbox{\O} &      &\vbl4 &{}_{n+N-1} \cr
 \multispan{14}\hblfil \cr
  {}_{n_{l+1}} &\vbl4 &    &\hbox{\O} &      &            &\vbl4 &{}_n &\vbl4
   &{}_{n+1} &{}_{\cdots}    &{}_{n+N-2} &\vbl4 &{}_{n+N-1} \cr}
 \right| = 0.\eqno(4.50)
$$
By the Laplace expansion, we have
$$
 \eqalignno{
    &|{}_{n_{l+1}},{}_{n+1},{}_{\cdots},{}_{n+N-2},{}_{n+N-1_{k+1}}|
|{}_n,{}_{n+1},{}_{\cdots},{}_{n+N-2},{}_{n+N-1}| \cr
  - &|{}_n,{}_{n+1},{}_{\cdots},{}_{n+N-2},{}_{n+N-1_{k+1}}|
|{}_{n_{l+1}},{}_{n+1},{}_{\cdots},{}_{n+N-2},{}_{n+N-1}| \cr
  + &|{}_{n+1},{}_{\cdots},{}_{n+N-2},{}_{n+N-1},{}_{n+N-1_{k+1}}|
|{}_{n_{l+1}},{}_n,{}_{n+1},{}_{\cdots},{}_{n+N-2}| = 0.
  &(4.51)}
$$
Substituting eqs.~(4.39)-(4.42) and (4.49) into eq.~(4.51), we obtain
$$\eqalignno{
& (1-ab)\tau_n(k+1,l+1)\tau_n(k,l) - \tau_n(k+1,l)\tau_n(k,l+1)\cr
& + a\tau_{n+1}(k+1,l)b\tau_{n-1}(k,l+1) = 0,&(4.52)\cr}
$$
which recovers eq.~(4.28).  This completes the proof that the Casorati
determinant is indeed the solution for D2DTL equation.  \par
\ \par
\noindent
{{\sect\uppercase\expandafter{\romannumeral5}. REDUCTION TO THE RT
EQUATION}\par\ \par
  Summarizing the arguments in {\sect\uppercase\expandafter{\romannumeral 4}}
, we see that the three bilinear equations,
$$
 D_xD_y\tau_n(k,l)\cdot\tau_n(k,l)
 = 2( \tau_n(k,l)\tau_n(k,l) - \tau_{n+1}(k,l)\tau_{n-1}(k,l) ),
 \eqno(5.1{\rm a})
$$
$$
 ( aD_x - 1 )\tau_n(k+1,l)\cdot\tau_{n-1}(k,l) + \tau_n(k,l)\tau_{n-1}(k+1,l)
 = 0,\eqno(5.1{\rm b})
$$
$$
\eqalignno{
& \tau_n(k+1,l+1)\tau_n(k,l) - \tau_n(k+1,l)\tau_n(k,l+1)\cr
& = ab( \tau_n(k+1,l+1)\tau_n(k,l) - \tau_{n+1}(k+1,l)\tau_{n-1}(k,l+1) ),
 &(5.1{\rm c})\cr}
$$
are simultaneously satisfied by the Casorati determinant,
$$
 \tau_n(k,l) = \left|\matrix{
  \varphi_1^{(n)}(k,l) &\varphi_1^{(n+1)}(k,l) &\cdots
   &\varphi_1^{(n+N-1)}(k,l) \cr
  \varphi_2^{(n)}(k,l) &\varphi_2^{(n+1)}(k,l) &\cdots
   &\varphi_2^{(n+N-1)}(k,l) \cr
  \vdots               &\vdots                 &
   &\vdots                 \cr
  \varphi_N^{(n)}(k,l) &\varphi_N^{(n+1)}(k,l) &\cdots
   &\varphi_N^{(n+N-1)}(k,l) \cr}
 \right|,\eqno(5.2)
$$
supposed that $\varphi_i^{(n)}$'s obey the dispersion relations,
$$
 \partial_x\varphi_i^{(n)} = \varphi_i^{(n+1)},\eqno(5.3{\rm a})
$$
$$
 \partial_y\varphi_i^{(n)} = \varphi_i^{(n-1)},\eqno(5.3{\rm b})
$$
$$
 \Delta_k\varphi_i^{(n)} = \varphi_i^{(n+1)},\eqno(5.3{\rm c})
$$
$$
 \Delta_l\varphi_i^{(n)} = \varphi_i^{(n-1)},\eqno(5.3{\rm d})
$$
where $\Delta_k$ and $\Delta_l$ are the same as those in eqs.~(4.31).
For example, we can take $\varphi_i^{(n)}$ as
$$
 \varphi_i^{(n)}(k,l)
 = p_i^n(1-p_ia)^{-k}(1-{1 \over p_i}b)^{-l}{\rm e}^{\eta_i}
 + q_i^n(1-q_ia)^{-k}(1-{1 \over q_i}b)^{-l}{\rm e}^{\xi_i},\eqno(5.4)
$$
$$
 \eta_i = {1 \over p_i}y + p_ix + \eta_{i0},\eqno(5.5{\rm a})
$$
$$
 \xi_i = {1 \over q_i}y + q_ix + \xi_{i0},\eqno(5.5{\rm b})
$$
where $p_i$, $q_i$ and $\eta_{i0}$, $\xi_{i0}$ are arbitrary constants
which correspond to the wave numbers and phase parameters of solitons,
respectively.  It can easily be seen that $\varphi_i^{(n)}$ in eq.~(5.4)
actually satisfies eqs.~(5.3).  \par
     The reduction is the technique to derive subhierarchies of equations
by restricting the solution space, in other words, imposing some conditions
to the solutions.  In the case of RT equation, we take in eq.~(5.4)
$$
 q_i = {1 \over p_i},\eqno(5.6)
$$
and
$$
 b = a.\eqno(5.7)
$$
On the conditions (5.6) and (5.7), $\varphi_i^{(n)}$ in eq.~(5.4) satisfies
$$
 (\partial_x+\partial_y)\varphi_i^{(n)} = (p_i+{1 \over p_i})\varphi_i^{(n)},
 \eqno(5.8{\rm a})
$$
$$
 \varphi_i^{(n)}(k-1,l-1) = (1-p_ia)(1-{1 \over p_i}a)\varphi_i^{(n)}(k,l).
 \eqno(5.8{\rm b})
$$
Hence we obtain
$$
 (\partial_x+\partial_y)\tau_n = S\tau_n,\eqno(5.9{\rm a})
$$
$$
 \tau_n(k-1,l-1) = P\tau_n(k,l),\eqno(5.9{\rm b})
$$
where
$$
 S = \sum_{i=1}^N (p_i+{1 \over p_i}),\eqno(5.10{\rm a})
$$
$$
 P = \prod_{i=1}^N (1-p_ia)(1-{1 \over p_i}a).\eqno(5.10{\rm b})
$$
By using eqs.~(5.9), eqs.~(5.1a) and (5.1c) are rewritten as
$$
 D_x^2\tau_n(k,l)\cdot\tau_n(k,l)
 = 2( \tau_{n+1}(k,l)\tau_{n-1}(k,l) - \tau_n(k,l)\tau_n(k,l) ),
 \eqno(5.11{\rm a})
$$
$$\eqalignno{
& \tau_n(k,l)\tau_n(k,l) - \tau_n(k+1,l)\tau_n(k-1,l)\cr
& = a^2( \tau_n(k,l)\tau_n(k,l) - \tau_{n+1}(k+1,l)\tau_{n-1}(k-1,l) ).
 &(5.11{\rm b})\cr}
$$
Here we may drop $y$ and $l$-dependence.  Thus eqs.~(5.11) and (5.1b) give
$$
 D_x^2\tau_n(k)\cdot\tau_n(k)
 = 2( \tau_{n+1}(k)\tau_{n-1}(k) - \tau_n(k)\tau_n(k) ),\eqno(5.12{\rm a})
$$
$$
 ( aD_x - 1 )\tau_n(k)\cdot\tau_{n-1}(k-1) + \tau_n(k-1)\tau_{n-1}(k)
 = 0,\eqno(5.12{\rm b})
$$
$$
 \tau_n(k+1)\tau_n(k-1) - \tau_n(k)\tau_n(k)
 = a^2( \tau_{n+1}(k+1)\tau_{n-1}(k-1) - \tau_n(k)\tau_n(k) ).
 \eqno(5.12{\rm c})
$$
Equations (5.12a) and (5.12c) are nothing but the bilinear form of the
one-dimensional TL and DTL equations.  From eqs.~(5.2) and (5.4)-(5.7),
the Casorati determinant solution of eqs.~(5.12) is
$$
 \tau_n(k) = \left|\matrix{
  \varphi_1^{(n)}(k) &\varphi_1^{(n+1)}(k) &\cdots &\varphi_1^{(n+N-1)}(k) \cr
  \varphi_2^{(n)}(k) &\varphi_2^{(n+1)}(k) &\cdots &\varphi_2^{(n+N-1)}(k) \cr
  \vdots             &\vdots               &       &\vdots                 \cr
  \varphi_N^{(n)}(k) &\varphi_N^{(n+1)}(k) &\cdots &\varphi_N^{(n+N-1)}(k) \cr}
 \right|,\eqno(5.13)
$$
$$
 \varphi_i^{(n)}(k)
 = p_i^n(1-p_ia)^{-k}{\rm e}^{\eta_i}
 + ({1 \over p_i})^n(1-{1 \over p_i}a)^{-k}{\rm e}^{\xi_i},\eqno(5.14)
$$
$$
 \eta_i = p_ix + \eta_{i0},\eqno(5.15{\rm a})
$$
$$
 \xi_i = {1 \over p_i}x + \xi_{i0}.\eqno(5.15{\rm b})
$$
\par
     Letting
$$
 \eqalignno{
  &f_n = \tau_n(n),&(5.16{\rm a}) \cr
  &g_n = \tau_{n-1}(n),&(5.16{\rm b}) \cr
  &{\bar g}_n = \tau_{n+1}(n),&(5.16{\rm c})}
$$
eqs.~(5.12) with $k=n$ become the bilinear form of RT (2.1), and eq.~(5.13)
gives its solution (3.1).  Thus we have completed the proof that the
$N$-soliton solution for the RT equation is given by the Casorati determinant
in eqs.~(3.1).  We note that besides the soliton solution, other solutions
such as the rational exponential soliton can be obtained in the same
way.  \par
\ \par
\noindent
{{\sect\uppercase\expandafter{\romannumeral6}. CONCLUDING REMARKS}\par
\ \par
     We have shown that the solution of the RT equation is given by the
Casorati determinant.  The composition of three Toda systems derives the
relativistic modification of the TL equation.  This fact means that the
Toda equation has quite fruitful content.  \par
     The difference between the relativistic and non-relativistic TL's is
only the choice of dependent variables in the D2DTL system.  In the
two-dimensional space of discrete independent variables $n$ and $k$, we
have the Casorati determinant solution $\tau_n(k)$.  If we take $\tau_n(n)$
as the dependent variable $f_n$, we obtain the relativistic version of TL
through the variable transformation (2.2a).  The difference interval $a$
gives the light speed ${\rm c}$.  On the other hand, adopting $\tau_n(k)$
with fixed $k$ as the dependent variable, we get the original TL in the
same variable transformation.  Namely, the RT is a sublattice of the D2DTL
while the TL is another sublattice.  \par
     It is well known that the other soliton equations such as the KdV,
modified KdV, sine-Gordon and nonlinear Schr\"odinger equations are closely
related to or derived from the Toda equation by suitable limiting
procedures.  Therefore we expect that the relativistic version of
these
equations can also be derived from the RT equation.  We conjecture
how to
construct such relativistic soliton equations and their solutions.
\item{(1)} For a given soliton equation, consider its integrable
           discretization.
\item{(2)} Derive the proper two-dimensional lattice system
           (before reduction).
\item{(3)} Take the different sublattice, that is, the different
direction
           of propagation.
\item{(4)} In a suitable variable transformation, the relativistic
equation
           corresponding to the original one will be obtained.  \par
\noindent
It may be interesting to investigate the relativistic version of
nonlinear integrable equations in this scenario.  \par
     It should be noted that the RT equation has many other aspects and
there are several types of determinant solutions besides the Casorati one.
Actually the Wronskian of two-directional type yields quite different
solutions from the present $N$-soliton solution.  We can derive some
interesting solutions from the Wronskian, which will be reported
elsewhere.  \par
\ \par
\noindent
{\sect ACKNOWLEDGMENT}\par
\ \par
     The authors are grateful to Professor J. Hietarinta and
Professor R. Hirota for valuable discussions.  \par
\ \par
\ \par
\item{${}^1$} S. N. M. Ruijsenaars, Commun. Math. Phys. {\bf 133}, 217~(1990).
\item{${}^2$} M. Bruschi and O. Ragnisco, Phys. Lett. A {\bf 129}, 21~(1988);
{\bf 134}, 365(1989).
\item{${}^3$} Yu. B. Suris, Phys. Lett. A {\bf 145}, 113~(1990).
\item{${}^4$} R. Hirota, {\it Solitons}, ed. by R. K. Bullough and P. J.
          Caudrey (Springer, Berlin, 1980).
\item{${}^5$} J. Satsuma, J. Phys. Soc. Jpn. {\bf 46}, 359~(1979).
\item{${}^6$} N. C. Freeman and J. J. C. Nimmo, Phys. Lett. A {\bf 95},
1~(1983); J. J. C. Nimmo, Phys. Lett. A {\bf 99}, 281~(1983); N. C. Freeman,
IMA J. Appl. Math. {\bf 32}, 125~(1984).
\item{${}^7$} R. Hirota, M. Ito and F. Kako, Prog. Theor. Phys. Suppl.
No. 94, 42~(1988).
\item{${}^8$} M. Sato, RIMS Kokyuroku 439, 30~(1981).
\item{${}^{9}$} M. Sato and Y. Sato, {\it Nonlinear Partial Differential
Equations in Applied Science}, ed. H. Fujita, P. D. Lax and G. Strang
          (Kinokuniya/North-Holland, Tokyo, 1983), p.259.
\item{${}^{10}$} E. Date, M. Kashiwara, M. Jimbo and T. Miwa, {\it Non-linear
           Integrable Systems --- Classical Theory and Quantum Theory}, ed.
           M. Jimbo and T. Miwa (World Scientific, Singapore, 1983), p.39.
\item{${}^{11}$} M. Jimbo and T. Miwa, Publ. RIMS, Kyoto Univ. {\bf 19}, 943~
(1983).
\item{${}^{12}$} T. Miwa, Proc. Jpn. Acad. {\bf 58A}, 9~(1982).
\item{${}^{13}$} K. Ueno and K. Takasaki, {\it Group Representation and
Systems of Differential Equations}, Adv. Stud. in Pure Math. {\bf 4}
           (Kinokuniya, Tokyo, 1984), p.1.
\item{${}^{14}$} J. Hietarinta and J. Satsuma, Phys. Lett. A {\bf 161}, 267~
(1991).
\item{${}^{15}$} J. Matsukidaira, J. Satsuma and W. Strampp, Phys. Lett. A
{\bf 147}, 467~(1990); J. Matsukidaira and J. Satsuma, J. Phys.
Soc. Jpn. {\bf 59}, 3413~(1990); Phys. Lett. A {\bf 154}, 366~
(1991); J. Hietarinta, K. Kajiwara, J. Matsukidaira and J. Satsuma,
           {\it Nonlinear Evolution Equations and Dynamical Systems},
           ed. by M.~Boiti, L.~Martina and F.~Penpinelli (World Scientific,
          Singapore, 1992).
\item{${}^{16}$} Y. Ohta, R. Hirota, S. Tsujimoto and T. Imai, to be published
 in J. Phys. Soc. Jpn.
\item{${}^{17}$} R. Hirota, Y. Ohta and J. Satsuma, Prog. Theor. Phys. Suppl.
No. 94, 59~(1988).
\item{${}^{18}$} R. Hirota and J. Satsuma, Prog. Theor. Phys. Suppl. No. 59,
64(1976).
\item{${}^{19}$} R. Hirota, J. Phys. Soc. Jpn. {\bf 50}, 3785~(1981).
\end